# Capacitance-voltage profiling techniques for characterization of semiconductor materials and devices

Miron J. Cristea

*'Politehnica' University of Bucharest, 313 Spl. Independentei, Bucharest, Romania; E-mail: miron.cristea@gmail.com*

**Abstract** This work re-defines the well-known C-V (capacitance-voltage) measurement technique, in the view of a new physics formula, discovered in 2006 [1].

## 1. Introduction

In 1960, J. Hilibrand and R. D. Gold developed a formula for capacitance-voltage measurement of the impurity/doping profiles of semiconductor junctions [2]. Their formula was later included in reference books of A.S. Grove [3], Simon Sze [4] and others. The method involves the measurement of the junction barrier capacitance at different reverse voltage biases. From C(V) data then N(x) is inferred – N being the doping concentration, usually measured in $cm^{-3}$ and x is the spatial coordinate (cm or µm).

## 2. Advantages of the C-V technique

Compared with other semiconductor profiling techniques (spreading resistance, differential conductance, Hall effect, SIMS, RBS etc.), the C-V method is an electric, non-destructive measurement of the barrier capacitance of semiconductor junctions, like p-n junctions, metal- semiconductor junctions and even metal-oxide-semiconductor (MOS) structures. This non-destructive character and large applicability gave the method a widespread, almost universal usage in the semiconductor industry.

## 3. The theory of Hilibrand and Gold C-V formula

### 3.1 Doping profile and electric charge density

In order to obtain the doping profile, the C-V (capacitance versus voltage) measurement is done using an asymmetric semiconductor junction, for example a $p^+n$ junction (Figure 1). One can see in Fig.1 that when the concentrations of acceptors from the $p^+$ side equals the concentration of donors from the n side, the junction is formed at $Xj = 0$ coordinate.

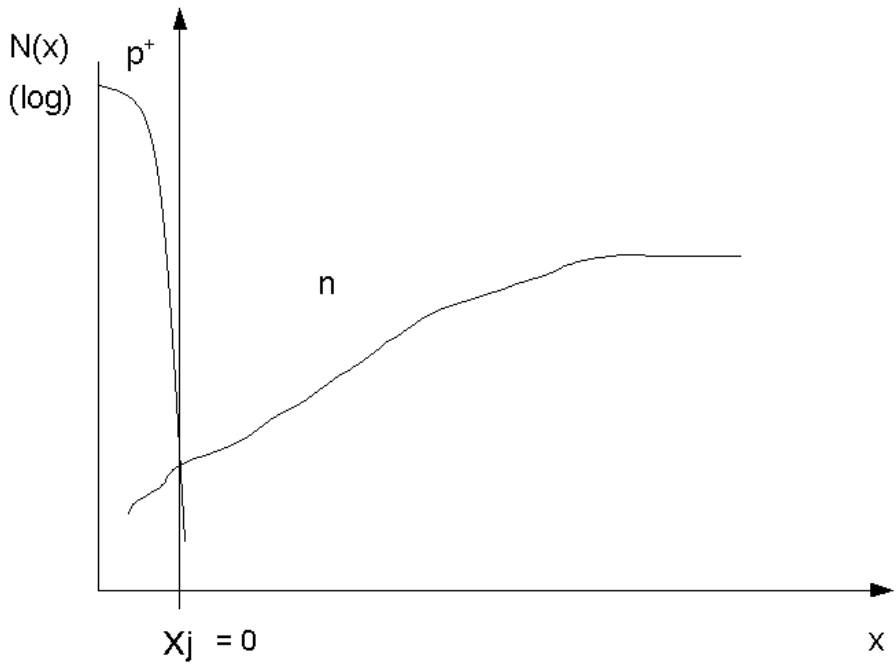

**Fig. 1.** Doping concentrations in an asymmetric p⁺n junction

When such a junction is reverse-biased, the space-charge region (SCR) spreads on both sides of the junction, but mainly in its low doped side (n). In the depletion-approximation (used by Hilibrand and Gold), the electric charge distribution looks like in Figure 2. Neglecting the spreading of the space-charge region in the heavily doped side of the junction, then W is equal to the SCR width. The positive electric charge (of the doping donors) in the n side of the junction is precisely balanced by the negative charge (of the acceptors) in the p⁺ side. In order to find the doping concentration at the current coordinate (W in Fig.2), the barrier capacitance is measured, knowing that:

$$C = \varepsilon / W \qquad (1)$$

where $\varepsilon$ is the permittivity of the semiconductor material (C expressed in F/cm$^{-2}$, usually).
Also, since

$$C = dQ/dV \qquad (2)$$

and (Fig.2.)

$$dQ = qN(W)dW \qquad (3)$$

then from (1)

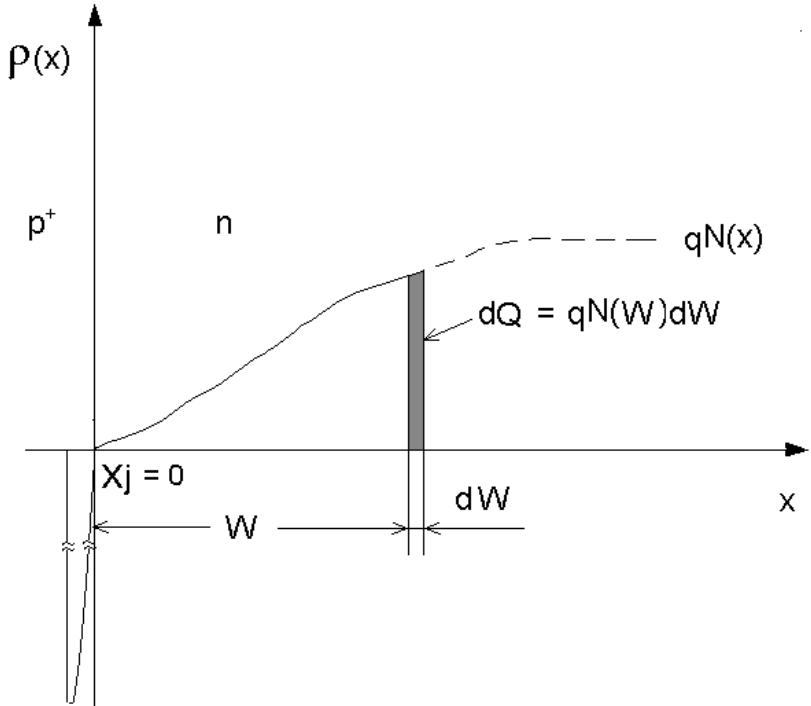

**Fig. 2.** Electric charge distribution in an asymmetric p⁺n junction (depletion approximation)

$$dW = d(\varepsilon C^{-1}) = -\varepsilon C^{-2} dC \quad (4)$$

and from (2) and (3)

$$dV = dQ/C = -qN(W)\,\varepsilon C^{-2} dC/C \quad (5)$$

Results the doping concentration as:

$$N(W) = -(C^3/q\varepsilon)/(dC/dV) \quad (6)$$

or

$$N(W) = (2/q\varepsilon)/[d(1/C^2)/dV] \quad (7)$$

which is the most used form, deduced by Hilibrand and Gold.

An example of measuring the bulk doping concentration of a silicon sample is given in Figure 3. Calculating the slope of $(1/C^2)$ versus voltage graph, the bulk concentration $N_B$ is found, since for a uniformly doped semiconductor [3]:

$$1/C^2 = 2(V_R + V_{bi})/q\varepsilon N_B \tag{8}$$

Also, the value of the built-in voltage $V_{bi}$ of the junction is found at the intersection of the ($1/C^2$) line with the horizontal axis. From figure 3 and eq.(8), $1/C^2 = 0$ when $V = -V_R = V_{bi}$.

In fact, for a more accurate measuring, $2kT/q$ (about 50 mV) should be added to the obtained value of $V_{bi}$ [5] (k is the Boltzmann constant and T the absolute temperature), but since $V_{bi}$ is around 700 mV this correction is negligible.

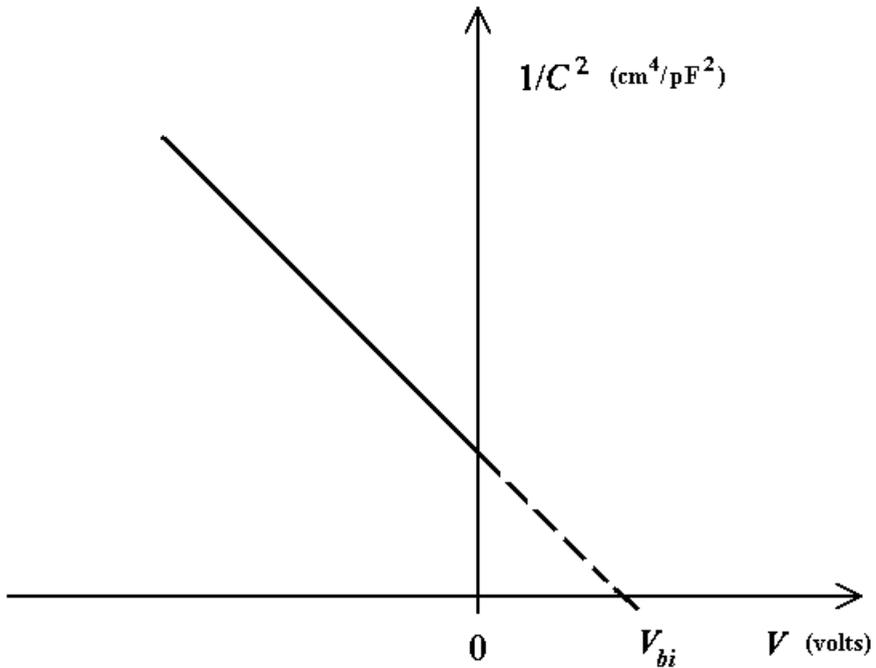

**Fig. 3.** $1/C^2$ plot of the barrier capacitance used to determine the background concentration $N_B$ and built-in potential $V_{bi}$ of a semiconductor junction

### 3.2 Flaws of the classic formula

Over the years, researchers in the semiconductor industry found discrepancies between the C-V formula results and actual doping profiles and they developed various reverse engineering and iterative methods to overcome them [6,7].

There were also observed limitations very close to the metallurgical junction and especially for abrupt and linear junctions.

The inaccuracies observed near x = 0 are due to the fact that, when the doping profile abruptly changes in a scale smaller than the Debye length $L_D$, its variation cannot be resolved and the analysis is no longer valid [5].

The main discrepancies are caused by the limited validity of the depletion approximation at the edge of the depletion zone. Actually, in formula (3) appears the **net** concentration N(W)-n(W) and not the doping concentration N(W), where n(W) is the mobile carrier concentration (electrons), neglected in the depletion approximation (Figure 4). From Fig.4 it is clear that the targeted doping concentration N(W) is 2-3 times or even higher than the measured value N(W)-n(W).

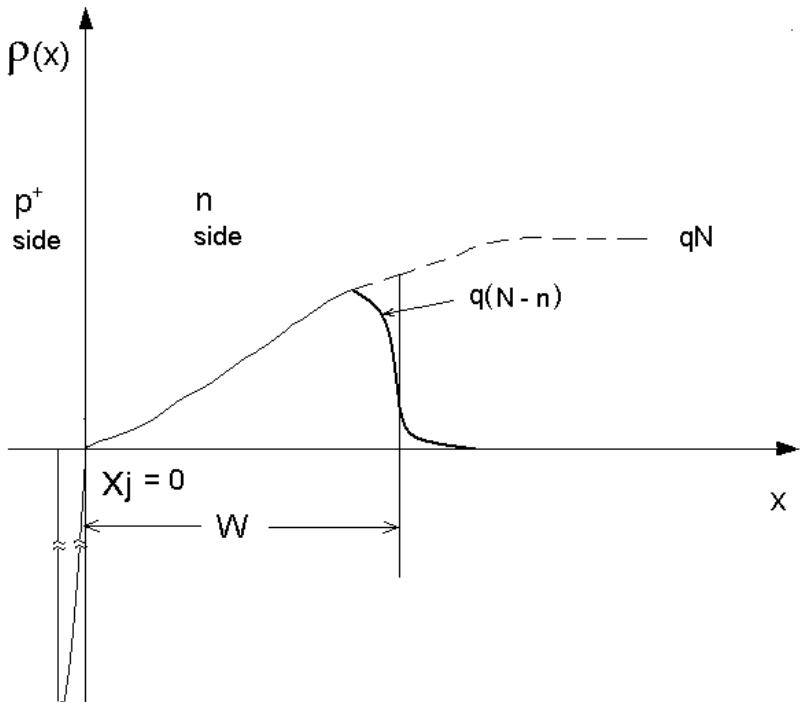

**Fig. 4.** Electric charge distribution in an asymmetric p$^+$n junction – the real case

## 4. Deduction of a new, correct formula

### 4.1 A new formula of physics

In this paragraph a new formula will be deduced from the Gauss's Law and the electric field (E)/electric potential (V) relation. From these two formulas was deduced also the well-known Poisson equation. Nevertheless, this will be an entirely new formula, able of solving problems that Poisson equation couldn't solve, as shown below.

The Gauss's Law (below in differential form) connects the electric field with the electric charge density.

$$dE/dx = \rho(x)/\varepsilon \tag{9}$$

This holds for any linear material or space region where $\varepsilon$ does not depend on the electric field intensity.

The electric field is also connected with the electric potential V along the field lines, according to:

$$E = -dV/dx \tag{10}$$

By writing (9) as

$$dE = \rho(x)dx/\varepsilon \tag{11}$$

multiplying this equation by x and taking into account that

$$d(xE) = xdE + Edx \tag{12}$$

the next is obtained:

$$d(xE) - Edx = x\rho(x)dx/\varepsilon \tag{13}$$

The integration of (13) over the space charge region defined between coordinates X1 and X2 gives

$$\int_{X1}^{X2} \frac{x\rho(x)}{\varepsilon} dx = \int_{X1}^{X2} d(xE) - \int_{X1}^{X2} Edx \tag{14}$$

and taking (10) into account, then

$$\int_{X1}^{X2} \frac{x\rho(x)}{\varepsilon} dx = \int_{X1}^{X2} d(xE) + \int_{X1}^{X2} dV \tag{15}$$

Both terms in the right hand of this equation are perfect integrals. So we can integrate the right side like:

$$\int_{X1}^{X2} \frac{x\rho(x)}{\varepsilon} dx = V(x_2) - V(x_1) + x_2 E(x_2) - x_1 E(x_1) \tag{16}$$

The applicability of this new equation is not limited to semiconductor junctions, but it extends in the electro-magnetic field theory [1].

### 4.2. Application to semiconductor junctions

Since the electric field is zero at both ends of the SCR [8], particularization of (16) to semiconductor junctions leads to

$$\int_{SCR} \frac{x\rho(x)}{\varepsilon} dx = V_{bi} - V_F \qquad (17)$$

where $V_{bi}$ is the built-in voltage of the junction and $V_F$ is the externally applied forward bias. If the junction is subjected to reverse bias, $V_F$ should be replaced with $-V_R$, therefore equation (17) becomes:

$$\int_{SCR} \frac{x\rho(x)}{\varepsilon} dx = V_{bi} + V_R \qquad (18)$$

In the case of homogenous semiconductor junctions, this formula can be written as:

$$\frac{1}{\varepsilon} \int_{SCR} x\rho(x) dx = V_{bi} + V_R \qquad (19)$$

since the permittivity is constant throughout the material.

However, in the case of hetero-junctions or other types of junctions in which more than one material is encountered, the following form of equation (18) should be applied:

$$\int_{SCR1} \frac{x\rho(x)}{\varepsilon_1} dx + \int_{SCR2} \frac{x\rho(x)}{\varepsilon_2} dx + \ldots + \int_{SCRn} \frac{x\rho(x)}{\varepsilon_n} dx = V_{bi} + V_R \qquad (20)$$

where SCR1, SCR2... SCRn are the fractions of the overall space charge region corresponding to the *n* semiconductor materials used for the junction fabrication. Another form of (19) can also be used when the particular geometry of the device requires:

$$\frac{1}{\varepsilon} \int_{SCR} (x + k_x) \rho(x) dx = V_{bi} + V_R \qquad (21)$$

where $k_x$ is a constant distance. This equation can be deduced following the algorithm used for (16), or by noticing that

$$\frac{k_x}{\varepsilon} \int_{SCR} \rho(x)\, dx = 0 \qquad (22)$$

due to the space charge equilibrium law for the electric charge on both sides of the junction.

## 5. The new C-V extraction technique

### 5.1 Practical example

In Figure 5 is depicted the doping profile of the base-emitter junction of a bipolar transistor showing the extension of the space charge region (SCR) occurring mostly in the lightly-doped side of the junction (base).

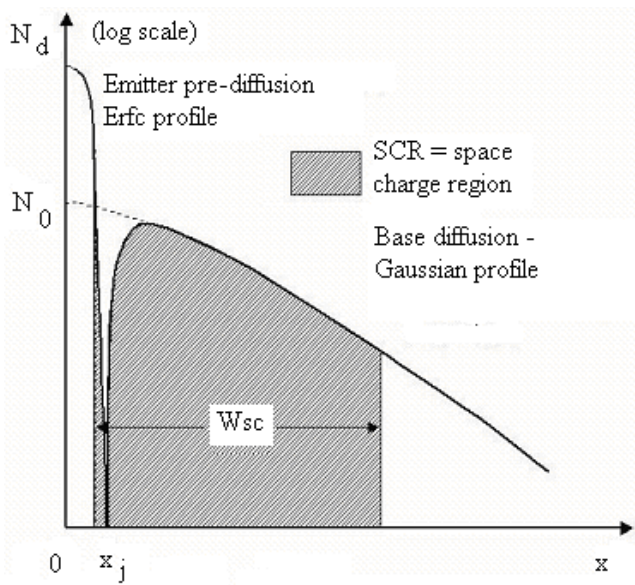

**Fig. 5.** Base-emitter junction doping profile and space charge region (SCR) of a bipolar transistor with diffused base

Supposing that the base-emitter junction is reverse-biased, we obtain by the integration of equation (18) the formula of the barrier capacitance of the junction [1]:

$$C_b = \cfrac{\varepsilon}{L_d \left( \sqrt{-\ln\left[ e^{-\frac{x_j^2}{L_d^2}} - \frac{2\varepsilon}{qN_0 L_d^2}\left(V_R + V_{bi}\right) \right]} \right) - x_j} \qquad (23)$$

Above, $x_j$ is the junction depth, $N_0$ is the surface concentration of the Gaussian diffusion and Ld is the technological diffusion length of the doping impurities $L_d = 2\sqrt{D_i t_d}$, with $D_i$ the diffusion constant of the impurities and $t_d$ their diffusion time during the fabrication of the p-n junction.

**5.2 The shallow semiconductor junction barrier capacitance and SCR width**

For shallow junctions, $x_j$ is negligible compared to $W_{SC}$ like in Fig. 5, therefore the barrier capacitance can be written as:

$$C_b \simeq \frac{\varepsilon}{L_d} \left\{ \sqrt{-\ln\left[ e^{-\frac{x_j^2}{L_d^2}} - \frac{2\varepsilon}{qN_0 L_d^2}\left(V_R + V_{bi}\right) \right]} \right\}^{-1} \qquad (24)$$

For such junctions the SCR width is given by $W_{SC} = \varepsilon/C_b$:

$$W_{SC} = L_d \sqrt{-\ln\left[ e^{-\frac{x_j^2}{L_d^2}} - \frac{2\varepsilon}{qN_0 L_d^2}\left(V_R + V_{bi}\right) \right]} \qquad (25)$$

**5.3 The case of junctions with xj << Ld (deep diffused base junction)**

For certain types of junctions, the condition xj << Ld is fulfilled (note that this is a technological condition). Since Ld is large compared to xj, we can say that this is a deep diffused base junction. For such junction, the exponential term in (24) and (25) can be taken as unity. The following simplified formulas are obtained:

$$W_{SC} = L_d \sqrt{-\ln\left[ 1 - \frac{2\varepsilon}{qN_0 L_d^2}\left(V_R + V_{bi}\right) \right]} \qquad (26)$$

$$C_b \simeq \frac{\varepsilon}{L_d} \left\{ \sqrt{-\ln\left[1 - \frac{2\varepsilon}{qN_0 L_d^2}(V_R + V_{bi})\right]} \right\}^{-1} \tag{27}$$

## 6. Experimental results

Semiconductor devices such as bipolar transistors, thyristors, IGBTs and other devices with diffused base share the same doping profile as depicted in figure 5 [1,9]. For these devices, and also for p-n diodes, varicap diodes and other semiconductor devices, it is of interest to find the parameters of the diffused base using a non-invasive, non-destructive technique. Such a technique is the well-known C-V measurement of the barrier capacitance of the p-n junction under varying reverse bias conditions. At various reverse voltage values, the barrier capacitance

$$C_B = A_J C_b \tag{28}$$

was measured [10]. Here $C_b$ is the specific capacitance and $A_J$ is the junction area.

The advantages of this technique are:

- it is non-destructive and measurements are done directly to the device terminals;
- the measurements can be performed with ordinary equipment;
- the parameters can be extracted with a small amount of data processing.

The measurements were done with a KEITHLEY C-V ANALYZER 590 Semiconductor Characterisation System using its sine voltage 15 mV rms test signal. Testing semiconductor devices with larger test signals can cause curve shape distortion and loss of detail. The frequency of 100kHz was used for the C-V measurements because it has better accuracy then 1MHz test frequency (0.12% vs. 0.29%) and reduced errors due to cabling or device series resistance. A 1MHz test frequency is traditionally specified in C-V test procedures requiring "high-frequency" device characteristics. The measurement circuit is very simple using the transistor connected with the emitter-base terminals at the measurement input and the collector decupled with 100nF capacitor to the ground of the C-V analyser.

The measured transistors were low power and medium power transistors. The low power transistors where BF214 and 2N2906 types. BF214 is a npn transistor designed for AM radio receivers and IF amplifier stages, with Vce=30V, Ic=30mA, Ptot=0.16W and fT=250MHz. 2N2906 is pnp transistor for general purpose small signal, with Vce=40V, Ic=0.6A and Ptot=1.8W.

Figure 6 presents with continuous line the experimental results of measuring the barrier capacitance of the reverse biased base-emitter junction of both transistors compared with calculated theoretical dashed curves using the extracted parameters.

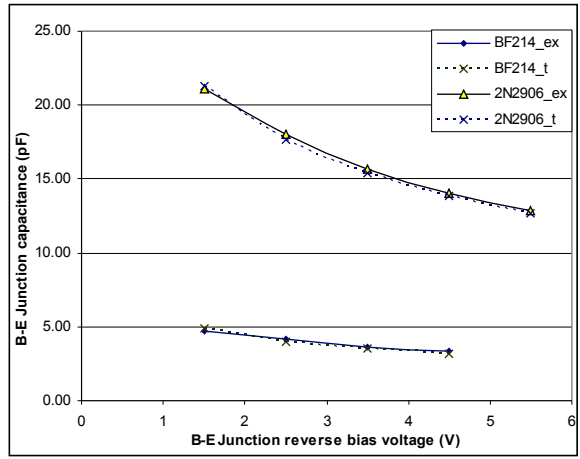

**Fig. 6.** The experimental (_ex) and theoretical (_t) data for junction capacity dependency function of emitter junction reverse bias voltage, for low power transistors.

From equations (27) and (28), the parameters of two bipolar transistor types were extracted with the aid of a curve-fitting program like EasyPlot or MathCAD. The extracted parameters are: $A_J$ – the junction area, $N_0$ – the surface concentration of the diffusion, $L_d$ – the technological diffusion length and $V_{bi}$ – the built-in junction potential. The values of the extracted parameters are presented in Table 1.

|  | $A_J$ [mm$^2$] | $L_d$ [µm] | $V_{bi}$ [V] | $N_0$ [cm-3] |
|---|---|---|---|---|
| BF214 (n-p-n) | $2.5\times10^{-3}$ | 0.6 | 0.7 | $10^{18}$ |
| 2N2906 (p-n-p) | $1.1\times10^{-2}$ | 0.5 | 0.7 | $10^{18}$ |

**Table 1.** Extracted parameters of low power transistors base-emitter junction

The medium power transistors were 2N2890 and BD237. 2N2890 is npn switching transistor with Vce=80V, Ic=3A and Ptot=5W. BD237 is npn type designed to be used in LF applications as drivers and in output stages, with Vce=80V, Ic=2A and Ptot=25W.
For these transistors, Fig. 7 presents with continuous line the experimental results of measuring the barrier capacitance of the reverse biased base-emitter junction of the transistors comparing with calculated theoretical dashed curves.

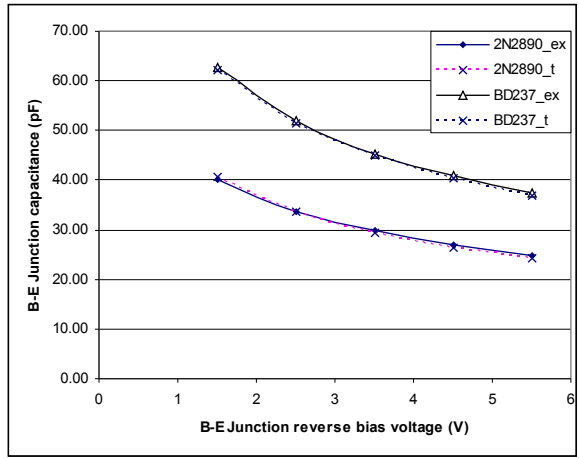

**Fig. 7.** The experimental (_ex) and theoretical (_t) data for junction capacity dependency function of emitter junction reverse bias voltage, for low power transistors

Using the same procedure the next parameters presented in Table 2 where obtained.

|  | $A_J$ [mm$^2$] | $L_d$ [μm] | $V_{bi}$ [V] | $N_0$ [cm-3] |
|---|---|---|---|---|
| 2N2890 (n-p-n) | $2.1 \times 10^{-2}$ | 0.6 | 0.7 | $10^{18}$ |
| BD237 (n-p-n) | $3.2 \times 10^{-2}$ | 0.6 | 0.7 | $10^{18}$ |

**Table 2.** The extracted parameters of medium power transistors

The obtained parameters are in good agreement with the known device layouts and technological process data.

## 7. Conclusion

Although widely used for about 50 years, the C-V doping profiling formula of Hilibrand and Gold from 1960 proved to be flawed. The main flaws are the facts that it doesn't measure the doping but rather the net electric charge concentration at the space-charge region boundary and that it cannot follow steep variations of the doping profile.
Based on a new physics formula discovered in 2006, a new C-V parameter extraction technique was established and applied to semiconductor junctions and devices. Once a suitable analytical model is established for the doping profile, this new method extracts the parameters of that model by integration over the entire space-charge region; therefore it is not very much influenced by the errors occurring at the space-charge region boundaries.

## 8. References

[1] Cristea, M.J. (2007). Calculation of the Depletion Region Width and Barrier Capacitance of Diffused Semiconductor Junctions with Application to Reach-Through Breakdown Voltage of Semiconductor Devices with Diffused Base, *Proceedings of the International Semiconductor Conference CAS 2007 (an IEEE event), October 2007, 978-1-4244-0847-4,* Sinaia, Romania

[2] Hilibrand, J. & Gold, R.D. (1960). Determination of impurity distribution in junction diodes from capacitance-voltage measurements. *RCA Review*, 21, 245-52, RCA Laboratories, Princeton, NJ, June 1960, 0033-6831

[3] Grove, A.S. (1967). *Physics and technology of semiconductor devices*, Wiley, 0471329983, New York

[4] Tsai, J.C.C. (1983). Diffusion, In: *VLSI Technology*, Sze, S., 186-187, McGraw Hill, 0-07-062686-3, New York

[5] Sze, S.M. & Ng, Kwok K. (2007). *Physics of Semiconductor Devices*, third ed., Wiley, 978-0-47 1-1 4323-9, New York

[6] Kokorev, M.; Maleev N. & Pakhnin D. (2000). Inverse modelling for C-V profiling of modulated-doped semiconductor structures, *Technical Proceedings of the 2000 International Conference on Modeling and Simulation of Microsystems MSM 2000*, U.S. Grant hotel, March 27-29, 2000, 0-9666135-7-0, San Diego

[7] Ouwerling, G. (1990). Physical parameter extraction by inverse modelling: application to one- and two-dimensional doping profiling. *Solid-St. Electronics*, 33, 757-771, 1990, 0038-1101

[8] Sze, S.M. (1981). *Physics of Semiconductor Devices*, second ed., Wiley, 0-471-09837-X, New York

[9] Mohan, N (2002). *Power Electronics: Converters, Applications, and Design*, 3rd Ed., John Wiley & Sons, 978-0471226932, New York

[10] Cristea, M.J. & Babarada, F. (2008). C-V Parameter Extraction Technique For Characterisation Of Diffused Junctions Of Semiconductor Devices, *Proceedings of the International Semiconductor Conference CAS 2008 (an IEEE event), October 2008, 978-1-4244-2004-9*, Sinaia, Romania